\documentclass[useAMS,usenatbib,usegraphicx]{mn2e}

\title{The old open cluster  NGC~2112: updated estimates of fundamental
  parameters based on  a membership analysis\thanks{This paper includes data gathered with the 6.5 Magellan Telescopes, located at Las Campanas Observatory, Chile}
\thanks{The data discussed in this paper will be made available at the WEBDA open cluster database {\tt http://www.univie.ac.at/webda}, which is maintained by E. Paunzen and
J.-C. Mermilliod}}
\author[Carraro  at al.]
{G. Carraro$^{1,2,3,4}$,  S. Villanova$^{2,5}$, P.
  Demarque$^{4}$, C. Moni Bidin$^{3}$, and M.V. McSwain$^{4,6}$
  \thanks{email: gcarraro@eso.org (GC), sandro.villanova@unipd.it (SV),
   demarque@astro.yale.edu(PD), mbidin@das.uchile.cl (CMB), mcswain@lehigh.edu(MVM)}\\
         $^1$ESO, Casilla 19001, Santiago 19, Chile\\
         $^2$Dipartimento di Astronomia, Universit\`a di Padova, vic. Osservatorio 3,
         Padova, Italy\\
         $^3$Departamento de Astronomia, Universidad de Chile, Casilla 36-D, Santiago, Chile\\
         $^4$Astronomy Department, Yale University, P.O. Box 208101 New Haven, CT 06520-8101 USA  \\
         $^5$Departamento de F\'isica, Facultad de Ciencias F\'isicas y Matem\'aticas
	 Universidad de Concepci\'on, Casilla 160-C, Concepci\'on, Chile\\
         $^6$Department of Physics, Lehigh University, 16 Memorial Drive East, Bethlehem, PA 18015, USA\\
         }

\begin{document}
\date{Accepted 1988 December 15. Received 1988 December 14; in original form 1988 October 11}
\maketitle

\label{firstpage}

\begin{abstract}
We report on a new, wide field ($20^{\prime} \times 20^{\prime}$),
multicolor ($UBVI$), photometric campaign
in the area of the nearby old open cluster NGC~2112. At the
same time, we provide medium-resolution spectroscopy of 35 (and high-resolution of additional 5)
Red Giant and Turn Off stars. This material is analyzed with the
aim to update the fundamental parameters of this traditionally
difficult cluster,
which is very sparse and suffers from heavy field star contamination.
Among the 40 stars with spectra, we identified 21 {\it bona fide}
radial velocity members which allow us to put more solid constraints on
the cluster's metal abundance, long suggested to be as low as the
metallicity of globulars.
As indicated
earlier by us on a purely photometric basis (Carraro et al. 2002), the cluster [Fe/H] abundance
is slightly super-solar ([Fe/H] =0.16$\pm$0.03) and close to the Hyades value, as inferred from a detailed abundance analysis 
of 3 of the 5 stars with higher resolution spectra.
Abundance ratios are also marginally super solar.\\
Based on this result, we revise the 
properties of NGC~2112 using stellar models from the
Padova and Yale-Yonsei groups. \\
For this metal abundance, we find the cluster's age,
reddening, and distance values are 1.8 Gyr, 0.60 mag,  and 940 pc, respectively.
Both the Yale-Yonsei and Padova models predict the same values
for the fundamental parameters within the errors.\\
Overall, NGC 2112 is a typical solar neighborhood, 
thin disk star cluster, sharing the same chemical properties
of F-G stars and open clusters close to the Sun.\\
This investigation
outlines the importance of a detailed membership analysis in the
study of disk star clusters.
\end{abstract}


%
\section{Introduction}
Gathering information on metal abundance and abundance ratios of many
Galactic clusters located in different region of the disk
and with different ages is mandatory to study the chemical evolution of the Galactic
disk. This, in turn, provides us with hints on the formation mechanism
of the disk and its relation with the other major
components of the Galaxy, the halo and bulge (Janes \& Phelps 1994, Carraro et al. 1998).\\
However, a frequent, well known problem in the study of open star clusters is
the stellar contamination from the general Galactic disk
field, which complicates the analysis of the Color Magnitude
Diagram (CMD).  It also makes it difficult to derive fundamental cluster
parameters, especially metallicity,  when only a few stars are observed.\\
\noindent
A notorious example in this context is the nearby old open
cluster NGC~2112 (Collinder 76, C~0551-0031, OCL~509), which
has a reputation of suffering from heavy field star contamination
(Brown et al. 1996), and for this reason its basic parameters
remain poorly constrained.\\
\noindent
The first investigation on NGC~2112 was
carried out by Richtler (1985, hereinafter R85), who obtained photographic BV
photometry for about 80 stars down to V= 15.
Although his photometry barely
reaches the cluster Turn-Off (TO), he nevertheless drew attention to
this probably old, so far neglected cluster, and he suggested that NGC 2112 has a
reddening of $\sim$ 0.5 mag and lies $\sim$800 pc from the Sun. By analyzing additional Stromgren photometry, he proposed that the
cluster had to be very metal poor ({\rm [Fe/H]} as low as -1.4).\\
A more accurate and deeper analysis was performed a few years later by Richtler \&
Kaluzny (1989). They obtained BV CCD photometry for about 500 stars in
a field of 200 arcmin$^2$. Additionally, they obtained moderate
resolution spectra for a handful of bright stars. Their conclusions
were that the cluster was very contaminated by field
stars. Nevertheless, they were able to strengthen the suggestions of
Richtler (1985, hereinafter R85) by confirming that the cluster is indeed old (3-5 Gyrs),
is located 700$\div$800 pc from the Sun and has a reddening of E(B-V)=0.60 mag.
Lacking a membership analysis, no further information on the cluster
metallicity was provided.\\
Photometry in the Washington system by Geisler(1987) and Geisler,
Claria \& Minniti (1991) seems to confirm the previous
suggestions that NGC 2112 is very metal poor ({\rm [Fe/H]} as low as -1.3). \\
\noindent
More recently, three spectroscopic campaigns have been carried out in the
field of the cluster.
Friel \& Janes (1993, hereinafter FJ93) present moderate resolution spectra of 6 stars.
Out of these, 5 are considered members, and an average $[Fe/H]$ of -0.52$\pm$0.21
has been found. This is significantly larger than all the previous
determinations.
Brown et al. (1996, hereinafter BWGO96) used the Blanco Echelle on CTIO to obtain high-resolution spectroscopy of 6 stars. They accepted  just one star as a definitive
member, providing an even higher metal
content value, {\rm [Fe/H]}=-0.15, only slightly lower than the solar
value.
Finally, Mermilliod \& Mayor (2007, hereinafter MM07) enlarged the sample of 
spectroscopic members to 3, out of 6 stars observed with Coravel.  They underlined
the need to obtain radial velocities for a larger sample down to V $\approx$
14.5 to better probe the shape of the Red Giant Branch (RGB).
At the same time, this would also provide much firmer metallicity estimates.\\
\noindent
In Carraro et al. (2002) we
reported a Jonhson BVI photometry
of the cluster down to magnitude V=20. Assuming the metal content
found by Brown et al. (1996), we found a reddening of
E(B-V)=0.63$\pm$0.14, a distance of 850$\pm$100 pc, and an age of
2.0$\pm$0.3 Gyr. We argued on a purely photometric basis that the metallicity
cannot be much lower than BWGO96 value, and we stressed the need for a new, more detailed spectroscopic investigation of the cluster.\\
\noindent
In this paper we attempt such an investigation by securing the deepest
and widest field coverage multicolor ($UBVI$) photometry of NGC~2112 to date.  At the same time, we provide moderate and high resolution spectroscopy of 40 stars.
With these data at hand, we present robust and updated determinations
of the cluster's basic properties.

\begin{figure}
\includegraphics[width=\columnwidth]{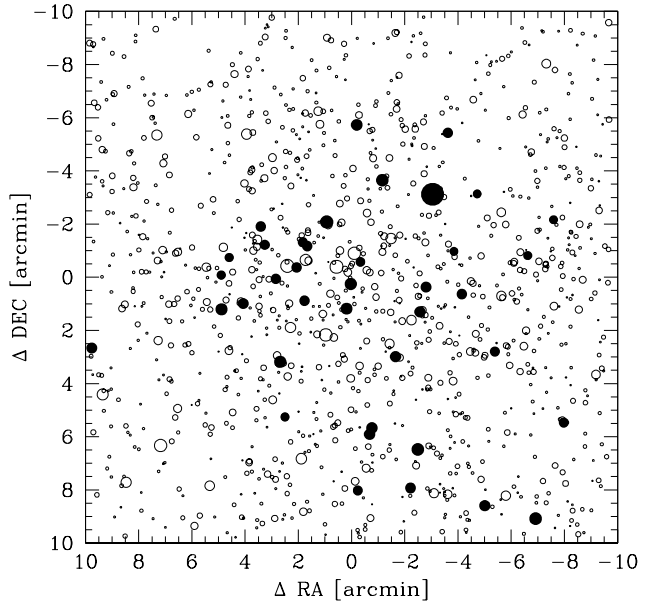}
\caption{V filter map of the field covered by our photometry. Stars (empty circles) are plotted
according to their magnitude.  North is up 
and East to the left.  The field is 20 arcmin on a side and centered at (X,Y) = (2046,2092) on star
$\#$593, which has $\alpha$ = 05:53:43.75 and $\delta$=   +00:23:56.4.
With filled circles we indicate stars observed spectroscopically.}
\label{mappa}
\end{figure}

%
\section{Observations and data reduction}
In this work we present photometry and spectroscopy in the field of NGC~2112
obtained with three different telescopes. For this reason,
the details of data acquisition and reduction are presented in the next
three sub-sections.

\subsection{Photometry}
$U,B,V,$ and $I$ images centered on NGC~2112  were obtained at the
Cerro Tololo Inter-American Observatory 1.0 m telescope, which is
operated by the SMARTS\footnote{http://www.astro.yale.edu/smarts/}
consortium. The telescope is equipped with a new 4k$\times$4k CCD
camera having a pixel scale of 0$^{\prime\prime}$.289/pixel, which
allows one to cover a field of $20^{\prime} \times 20^{\prime}$.
This allows us to cover the entire cluster,
which has an  estimated diameter of 18 arcmin (Dias et al. 2002).\\
Observations were carried out on November 30, 2005. Three Landolt
(1992) areas (TPhoenix, Rubin~149, and PG~0231+006) were also
observed to calibrate the instrumental magnitudes to the standard system.
The night was photometric with an average seeing of 1.1 arcsec. Data
were reduced using IRAF\footnote{IRAF is distributed by NOAO, which
is operated by AURA under cooperative agreement with the NSF.}
packages CCDRED, DAOPHOT, and PHOTCAL. Photometry was done employing
the point spread function (PSF) method (Stetson 1987). The covered
area is shown in Fig.~1, while Table~1  contains the
observational log.\\
The calibration equations read:

\begin{table}
\fontsize{8} {6pt}\selectfont
\caption{Journal of photometric observations of NGC~2112 and standard star
         fields together with calibration coefficients (November 30, 2005).}
\begin{tabular}{lcccccc}
\hline
\multicolumn{1}{c}{Field}         &
\multicolumn{1}{c}{Filter}        &
\multicolumn{1}{c}{Exposure time} &
\multicolumn{1}{c}{Seeing}        &
\multicolumn{1}{c}{Airmass}       \\
 &  \multicolumn{3}{c}{[sec.]} & [$\prime\prime$] & \\
\hline
\hline
NGC 2112        & U & 1200,60,5 & 1.1 & 1.150-1.280 \\
                & B &  900,30,3 & 1.0 & 1.150-1.280 \\
                & V &  600,30,1 & 1.0 & 1.150-1.280 \\
                & I &  600,30,1 & 1.0 & 1.150-1.280 \\
\hline
TPhoenix        & U & 180,200   & 1.0 & 1.024,1.444 \\
                & B &  90,120   & 1.1 & 1.023,1.447 \\
                & V &  20,30    & 1.1 & 1.024,1.450 \\
                & I &  40,40    & 1.1 & 1.022,1.452 \\
\hline
PG 0231+006     & U & 200,240   & 1.1 & 1.291,1.801 \\
                & B &  60,90    & 1.1 & 1.293,1.807 \\
                & V &  40,40    & 1.1 & 1.296,1.809 \\
                & I &  40,30    & 1.1 & 1.294,1.810 \\
\hline
Rubin~149       & U & 180,240   & 1.0 & 1.311,1.651 \\
                & B &  90,120   & 1.1 & 1.316,1.649 \\
                & V &  30,40    & 1.0 & 1.318,1.647 \\
                & I &  40,40    & 1.0 & 1.313,1.643 \\
\hline
\hline
Calibration     & \multicolumn {3}{l}{$u_1 = +3.285 \pm 0.004$} \\
coefficients    & \multicolumn {3}{l}{$u_2 = +0.032 \pm 0.006$} \\
                & \multicolumn {3}{l}{$u_3 = +0.46$}            \\
                & \multicolumn {3}{l}{$b_1 = +2.188 \pm 0.004$} \\
                & \multicolumn {3}{l}{$b_2 = -0.160 \pm 0.006$} \\
                & \multicolumn {3}{l}{$b_3 = +0.27$}            \\
                & \multicolumn {3}{l}{$v_{1bv} = +2.188 \pm 0.014$}\\
                & \multicolumn {3}{l}{$i_1 = +2.789 \pm 0.044$} \\
                & \multicolumn {3}{l}{$i_2 = +0.021 \pm 0.043$} \\
                & \multicolumn {3}{l}{$v_{2bv} = +0.017 \pm 0.018$}\\
                & \multicolumn {3}{l}{$i_3 = +0.06$}                \\
                & \multicolumn {3}{l}{$v_3 = +0.12$}                \\
                & \multicolumn {3}{l}{$v_{1vi} = +2.188 \pm 0.016$} \\
                & \multicolumn {3}{l}{$v_{2vi} = +0.013 \pm 0.016$} \\
\hline
\end{tabular}
\end{table}

          \begin{center}
           \begin{tabular}{lc}
            $u = U + u_1 + u_2 (U-B) + u_3 X$         & (1) \\
            $b = B + b_1 + b_2 (B-V) + b_3 X$         & (2) \\
            $v = V + v_{1bv} + v_{2bv} (B-V) + v_3 X$ & (3) \\
            $v = V + v_{1vi} + v_{2vi} (V-I) + v_3 X$ & (4) \\
            $i = I + i_1 + i_2 (V-I) + i_3 X$         & (5), \\
           \end{tabular}
          \end{center}

         \noindent
         where $UBVI$ are standard magnitudes, $ubvi$ are the instrumental
         magnitudes, $X$ is the airmass, and the derived coefficients are presented
         at the bottom of Table~1. To compute $V$ magnitudes when $B$ magnitudes
         were available, we used expression (3); otherwise we used
         expression (4). The standard stars in these fields provide
         a wide color coverage with
         $-1.217\leq (U-B) \leq 2.233$,$-0.298\leq (B-V) \leq 1.999$,
         and $-0.361\leq (V-I) \leq 2.268$. Aperture corrections were estimated
         in a sample of 15 bright stars and then applied to all stars. They
         amounted to 0.315, 0.300, 0.280, and 0.280 mag for the $U$, $B$, $V$,
         and $I$ filters, respectively.\\

 We cross-correlated our photometry with the photoelectric sequence of Richtler (1985) to check our zero-points. The cross-identifications are listed in Table~2,
         where for comparison purposes we approximate our values to two digits, as in R85.
         Values that have not been measured either by R85 or by us are replaced with 99.999.\\
         The differences in $V$, $B-V$ and $U-B$ between R85 and our study are illustrated in Fig.~2, and amount to:

\[ 
\Delta V = 0.02\pm0.06 ,
\]

\[ 
\Delta (B-V) = 0.02\pm0.04 , and
\]

\[ 
\Delta (U-B) = -0.01\pm0.09 
\]

\noindent
Our photometry is basically consistent with R85. Only $U-B$, although in agreement, exhibits
a significant scatter.\\

         The resulting CMDs are shown in Fig.~3 for three color combinations. 
	 The TO is located at V$\sim$14.5, (B-V) $\sim$ 1.1, and (V-I)  $\sim$ 1.3.

\begin{figure}
\includegraphics[width=\columnwidth]{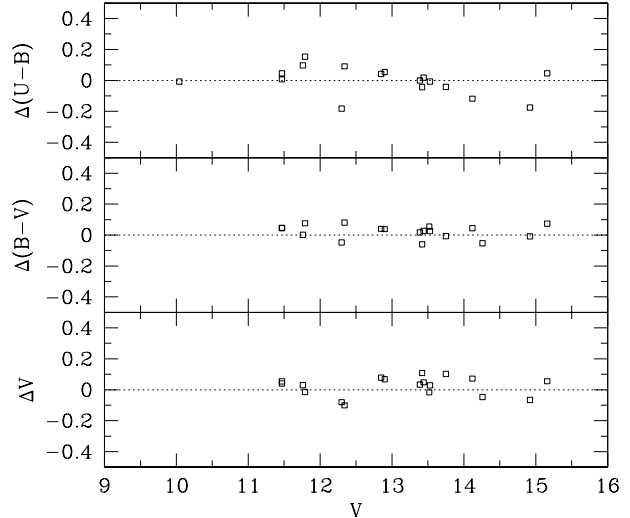}
\caption{Comparison of our photometry with the Richtler(1985, R85) photoelectric sequence.  }
\label{mappa}
\end{figure}

\begin{table*}
\caption{Comparison of the our photometry with the photoelectric sequence of Richtler(1985,R85)}
\begin{tabular}{lccccccc}
\hline
\multicolumn{1}{c}{$ID_{R85}$}         &
\multicolumn{1}{c}{$ID$}        &
\multicolumn{1}{c}{$V_{R85}$} &
\multicolumn{1}{c}{$(B-V)_{R85}$}        &
\multicolumn{1}{c}{$(U-B)_{R85}$}        &
\multicolumn{1}{c}{$V$}        &
\multicolumn{1}{c}{$(B-V)$}        &
\multicolumn{1}{c}{$(U-B)$}        \\
 \hline
1-01&	630&	11.47&	0.58&	0.38&   11.42&  0.54& 0.37\\    	
1-02&	593&	13.53&	1.04&	0.43&   13.50&  1.02& 0.43\\	
1-06&	585&	11.76&	0.72&	0.30&   11.73&  0.72& 0.20\\	
1-10&	512&	13.44&	1.12&	0.61&   13.39&  1.09& 0.59\\ 	
1-16&	382&	10.04&	2.25&	2.62&        &      & 2.62\\ 	
1-18&	401&	14.92&	0.99&	0.52&   14.99&  0.99& 0.70\\ 	
2-02&	601&	13.75&	1.00&	0.38&   13.65&  1.01& 0.42\\ 	
2-16&	422&	13.39&	1.66&	1.46&   13.36&  1.64& 1.46\\ 	
2-20&	724&	12.85&	0.85&	0.38&   12.77&  0.81& 0.34\\ 	
3-03&	749&	12.90&	0.95&	0.44&   12.83&  0.91& 0.39\\ 	
3-06&	909&	15.16&	1.11&	0.47&   15.11&  1.04& 0.42\\ 	
3-08&	919&	14.26&	1.74&	    &   14.31&  1.79& 1.75\\          
3-09&   1029&	14.12&	1.22&	0.60&   14.05&  1.17& 0.72\\ 	
3-10&	935&	13.52&	1.68&	    &   13.54&  1.62& 1.38\\        
3-16&   1123&	12.34&	2.11&	2.77&   12.44&  2.03& 2.68\\     	
3-17&   1026&	11.79&	2.15&	2.64&   11.80&  2.07& 2.49\\ 	
3-18&	954&	13.42&	1.49&	1.01&   13.31&  1.55& 1.05\\ 	
4-01&	711&	12.30&	1.51&	1.17&   12.38&  1.56& 1.35\\ 	
4-02&	759&	11.47&	1.47&	1.15&   11.43&  1.42& 1.10\\     	 
\hline
\end{tabular}
\end{table*}

\begin{table*}
\caption{Basic properties of the stars for which we secured spectroscopic observations. In the last column,
M indicates members according to radial velocity, NM indicates non-members}
\label{tab1}
\centering
\begin{tabular}{lrccrccccrrc}
\hline
ID & RA & DEC & U & B & V & $\sigma V$& I & ${\rm RV_{H}}$& $\mu_{\alpha} \cdot cos\delta$ & $\mu_{\delta}$ & \\
& $hh:mm:ss.s$ & $^{0}:^{\prime}:^{\prime\prime}$ & & & & & & $[km/sec]$ & [mas/yr]& [mas/yr] & \\
\hline
&&&&&&&&HYDRA observations&&&\\
\hline
 101&   05:53:11.7&  +00:18:43.4& 17.409& 15.983& 14.363& 0.017& 12.476&  42.47$\pm$0.68&   3.6$\pm$7.9& -10.8$\pm$8.2& NM\\
 124&   05:53:13.2&  +00:26:20.9& 17.397& 16.204& 14.873& 0.016& 13.226&  -3.16$\pm$0.78&   2.7$\pm$8.1& -11.0$\pm$7.8& NM\\
 167&   05:53:16.0&  +00:15:06.6& 16.433& 15.000& 13.313& 0.018& 11.315&  29.60$\pm$0.40&  -3.6$\pm$7.9&   8.8$\pm$7.8& M\\
 185&   05:53:17.1&  +00:25:00.3& 16.269& 15.834& 14.852& 0.015& 13.501&  36.53$\pm$0.96&  -3.6$\pm$7.8&  -3.2$\pm$7.8& NM\\
 244&   05:53:22.1&  +00:21:23.7& 15.913& 15.433& 14.407& 0.015& 13.026&  32.02$\pm$0.82&  -2.4$\pm$7.9&  -2.8$\pm$7.9& M\\
 274&   05:53:24.7&  +00:27:19.5& 16.907& 16.094& 14.909& 0.016& 13.247& -15.67$\pm$1.62&  -3.5$\pm$7.8&   4.2$\pm$7.9& NM\\
 323&   05:53:28.2&  +00:25:09.8& 16.715& 16.057& 14.896& 0.016& 13.305&  30.69$\pm$1.90&  -7.5$\pm$7.8&   3.0$\pm$7.9& M\\
 340&   05:53:29.1&  +00:29:37.5& 16.264& 15.418& 14.246& 0.016& 12.655&  32.27$\pm$0.74& -10.1$\pm$7.9&   2.9$\pm$7.9& M\\
 382&   05:53:31.4&  +00:27:18.7& 14.876& 12.249&       &      &       &  -6.18$\pm$0.22&  -1.2$\pm$1.0&  -7.5$\pm$1.1& NM\\
 399&   05:53:32.4&  +00:23:49.2& 15.600& 15.073& 13.998& 0.016& 12.518&  28.39$\pm$0.60&  -0.3$\pm$7.8&   5.5$\pm$7.8& M\\
 417&   05:53:33.3&  +00:22:53.3& 16.547& 15.369& 13.882& 0.017& 12.052&  29.49$\pm$0.25&              &              & M\\
 422&   05:53:33.7&  +00:17:42.9& 16.461& 15.000& 13.357& 0.018& 11.435& -29.32$\pm$0.26&  -6.3$\pm$7.8& -15.1$\pm$7.8& NM\\
 443&   05:53:34.8&  +00:16:16.0& 16.550& 15.505& 14.037& 0.016& 12.275&  29.52$\pm$0.34&  -5.9$\pm$7.8&   4.1$\pm$7.9& M\\
 478&   05:53:37.0&  +00:21:12.5& 15.528& 14.988& 13.898& 0.016& 12.428&  30.45$\pm$0.44&   1.2$\pm$7.8&   4.4$\pm$7.8& M\\
 512&   05:53:39.0&  +00:27:50.7& 15.077& 14.485& 13.392& 0.019& 11.805&  21.42$\pm$1.52&  -0.9$\pm$7.8&  -7.9$\pm$7.8& NM\\
 542&   05:53:40.9&  +00:18:17.1& 15.307& 14.820& 13.824& 0.015& 12.458&  30.78$\pm$0.54&  -1.8$\pm$7.9&   0.9$\pm$7.8& M\\
 566&   05:53:42.3&  +00:24:46.2& 16.006& 15.561& 14.539& 0.015& 13.141&  30.03$\pm$0.67&  -2.1$\pm$7.8&  -2.8$\pm$7.9& M\\
 577&   05:53:42.7&  +00:16:10.3& 16.016& 15.512& 14.454& 0.015& 13.036&  27.55$\pm$0.65&  -2.1$\pm$7.9&   5.1$\pm$8.0& NM\\
 580&   05:53:42.8&  +00:29:55.3& 15.317& 14.807& 13.739& 0.016& 12.380& -12.84$\pm$0.67&  10.8$\pm$7.9&   0.2$\pm$7.9& NM\\
 593&   05:53:43.7&  +00:23:56.5& 14.953& 14.517& 13.502& 0.017& 12.105&  44.78$\pm$0.49& -15.3$\pm$7.8&  -2.0$\pm$7.8& NM\\
 601&   05:53:44.4&  +00:23:00.7& 15.076& 14.654& 13.647& 0.019& 12.302&  30.02$\pm$0.60&   1.0$\pm$7.8&  -5.4$\pm$7.8& M\\
 655&   05:53:47.3&  +00:26:16.6& 16.416& 14.877& 13.155& 0.018& 11.036&  31.55$\pm$0.21&              &              & M\\
 656&   05:53:47.5&  +00:22:01.6& 15.337& 13.491& 11.724& 0.019&  9.663&  29.53$\pm$0.27&   3.2$\pm$1.7&  -0.8$\pm$1.7& M\\
 707&   05:53:50.3&  +00:25:22.1& 15.495& 15.148& 14.202& 0.015& 12.900&  30.81$\pm$0.81&  -9.2$\pm$8.0&  27.3$\pm$7.9& M\\
 714&   05:53:50.7&  +00:23:18.7& 15.672& 15.226& 14.227& 0.015& 12.843&  31.95$\pm$0.70&   1.9$\pm$7.8&   1.1$\pm$7.9& M\\
 737&   05:53:51.9&  +00:24:33.8& 15.526& 15.143& 14.157& 0.015& 12.825&  31.81$\pm$0.57&  -1.7$\pm$7.8&  -6.2$\pm$8.0& M\\
 770&   05:53:53.6&  +00:18:56.4& 16.235& 15.794& 14.767& 0.015& 13.379&  24.37$\pm$0.86&   1.7$\pm$7.8&  -5.8$\pm$7.9& NM\\
 782&   05:53:54.3&  +00:21:01.0& 15.466& 14.754& 13.510& 0.017& 11.883&  31.41$\pm$0.24&              &              & M\\
 794&   05:53:55.0&  +00:24:08.0& 15.727& 15.293& 14.280& 0.015& 12.884&  20.71$\pm$0.57&   3.0$\pm$7.8&   5.5$\pm$8.1& NM\\
 824&   05:53:56.6&  +00:25:24.8& 15.698& 15.327& 14.366& 0.015& 13.056&  30.65$\pm$0.58&  -0.9$\pm$7.8&   3.8$\pm$7.9& M\\
 890&   05:53:59.9&  +00:23:12.4& 15.393& 15.019& 14.051& 0.015& 12.728&  12.00$\pm$0.53&              &              & NM\\
 917&   05:54:02.0&  +00:24:56.4& 16.075& 15.724& 14.757& 0.016& 13.436&  22.70$\pm$1.16& -12.0$\pm$7.8&   0.2$\pm$8.2& NM\\
 935&   05:54:03.1&  +00:22:59.5& 16.546& 15.162& 13.537& 0.018& 11.505&  38.33$\pm$0.22&   1.0$\pm$7.8&  -4.3$\pm$7.8& NM\\
 936&   05:54:03.2&  +00:24:16.8& 15.987& 15.671& 14.719& 0.015& 13.384& -26.71$\pm$1.81& -12.9$\pm$7.8&   4.2$\pm$7.8& NM\\
1140&   05:54:22.6&  +00:21:32.6&       & 14.947& 13.964& 0.021& 12.581&   3.89$\pm$1.01&  -1.0$\pm$7.8&   3.3$\pm$7.9& NM\\
\hline
&&&&&&&&MIKE observations&&&\\
\hline
 261&   05:53:23.6&  +00:15:35.9&   17.100& 15.619& 13.955& 0.015& 11.987& 19.84$\pm$0.14&  -6.3$\pm$7.8&  -6.7$\pm$7.9& NM\\
 304&   05:53:27.1&  +00:23:33.3&   16.059& 15.402& 14.181& 0.015& 12.583& 29.27$\pm$0.15&  -4.2$\pm$7.8&  -0.6$\pm$7.9& M\\
 535&   05:53:40.6&  +00:18:31.4&   15.326& 14.826& 13.770& 0.017& 12.317& 29.55$\pm$0.25&   2.7$\pm$7.9&  -1.9$\pm$7.9& M\\
 717&   05:53:50.9&  +00:25:30.8&   15.676& 15.308& 14.336& 0.012& 12.998& 39.03$\pm$0.24&   4.1$\pm$7.9&   0.6$\pm$7.9& NM\\
 836&   05:53:57.3&  +00:26:06.1&   16.681& 15.622& 14.171& 0.014& 12.402& 29.26$\pm$0.16&  -6.1$\pm$8.1&   3.2$\pm$7.9& M\\
\hline
\end{tabular}
\end{table*}

\subsection{Spectroscopy: HYDRA observations}
Medium resolution spectroscopic observations were carried out on the night of 
2006 Feb 15 (Julian Date 2453783.57006) with
the Hydra spectrograph on-board the Wisconsin Indiana Yale NOAO (WIYN)
telescope at Kitt Peak National Observatory under photometric conditions and
typical seeing of 1.0 arcsec. The Multi-Object Spectrograph (MOS)
consists of the Hydra positioner, which in 20 minutes can place 89
fibers within the 1$^0$ diameter focal plane of the telescope to
0.2 arcsec precision. This project employed the 3 arcsec diameter
red-optimized fiber bundle.\\
The fibers feed a bench-mounted
spectrograph in a thermally isolated room. With the echelle grating
and the Bench Spectrograph Camera, the system produces a resolution of
15000 at 6560 \AA~. The wavelength coverage of 400 \AA\ around the central
wavelength of 6560 \AA\ provides a rich array of narrow absorption
lines. We observed 35 TO-RGB stars with 2$\times$45 minute
exposures, for a grand total of 1.5 hr of actual photon collection
time on each star.\\
The 35 stars were selected from the
photometric catalog presented in previous section. By using UCAC2 catalog (Zacharias et al 2004) as
reference, we converted pixel coordinates into 2000.0 equinox Right Ascension and Declination
using 50 stars as input. The astrometry precision is 0.3 arcsec.
The selected stars for Hydra are
candidate RGB and TO
stars according to their position in the CMD and have the right magnitudes to be observed with the WIYN 3.6 m
telescope. We restricted the sample to stars brighter than V$\sim$15.0.
The stars are listed in Table~2, where column (1) reports
numbering. In the following columns we report
2000.0 equinox coordinates, magnitude and colors, heliocentric radial velocity, and proper motion
components from UCAC2 (Zacharias et al. 2004). In the last column
an indication of membership is provided (see next section).
For some stars ($\#$782, 655, 890 and 417), no proper motions are
available from UCAC2, probably because these stars have close companions.

\begin{figure}
\includegraphics[width=\columnwidth]{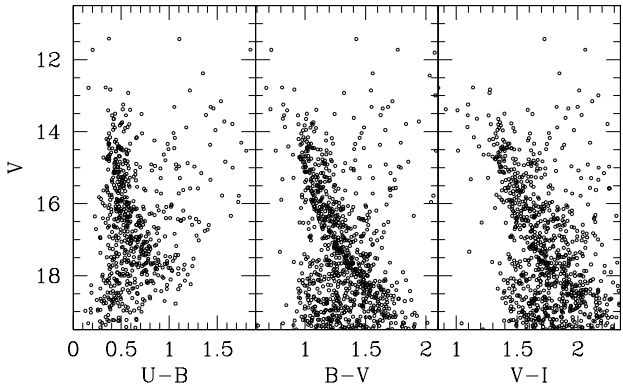}
\caption{CMD for the stars in the field of NGC 2112 in the V vs U-B (left panel),
         V vs B-V (medium panel) and V vs V-I (right panel).}
\label{fig1}
\end{figure}

Images were pre-reduced using IRAF \footnote{IRAF is distributed by the
National Optical Astronomy Observatory, which is operated by the
Association of Universities for Research in Astronomy, Inc., under
cooperative agreement with the National Science Foundation.}
including bias subtraction,
flat-field correction, frame combination, extraction of spectral
orders, wavelength calibration, sky subtraction, and spectral
normalization.
Some spectra turned out to have a very low signal-to-noise ratio (S/N),
although all the observed stars have practically the same
magnitude. This could happen for two reasons: the first is an
imperfect pointing of the fiber, and the second is possibly bad
fiber transmission.

\subsection{Spectroscopy: MIKE observations}
Echelle spectrograms of stars $\#$535, $\#$261,  $\#$717, $\#$304 and $\#$836 (see Table 2)
were obtained on 2007 October 29 with the Magellan
Inamori Kyocera Echelle (MIKE)
spectrograph mounted on the Nasmyth focus of Landon Clay 6.5m
telescope at the Magellan Observatory.
Data were obtained with both the blue and red arms.
The slit was 0.7 arcsec wide, which yielded a resolution
R=33000, and the CCD was
binned in steps of 2 pixels in the dispersion direction.  
The typical seeing was 0.6-0.8 arcsec.
We used quartz lamp
images without the diffuser in position for flat field correction, and
the wavelength calibration
was performed with  ThAr lamp images that were taken just before and
after the five stellar exposures.
The dark current was checked by examining several dark exposures and
was found to be insignificant.
The optimum algorithm (Horne 1986) was used to extract the spectra,
which were also sky-subtracted
and normalized using IRAF routines.
Additional details on the spectra are reported in Table~3.

\begin{table}
\fontsize{8} {6pt}\selectfont
\caption{Observational details of the 5 stars observed with MIKE}
\label{tab1}
\centering
\begin{tabular}{lccc}
\hline
ID & Julian Date & Exposure & S/N \\
& & sec & \\
\hline
 535&   2454303.82317&  500&    80\\
 261&   2454403.81600&  800&    60\\
 717&   2454403.83538& 1200&   100\\
 304&   2454403.85079& 1600&    90\\
 836&   2454403.87042& 1300&    80\\
\hline
\end{tabular}
\end{table}

\begin{table*}
\caption{Radial velocity: comparison with literature data}
\label{tab1}
\centering
\begin{tabular}{lccccccc}
\hline
R85 & ID& V & (B-V)& $RV_{H}$& BWGO96 & FJ93 & MM07\\
\hline
1-16 & 382 &       &      &  -6.18$\pm$0.22&   -3.6          &     & - 4.04\\
2-4  & 656 & 11.724& 1.767&  29.53$\pm$0.27&   30.04         &     &       \\
2-16 & 422 & 13.357& 1.643& -29.32$\pm$0.26&  -23.3          & -22 & -29.02\\
3-16 &1123 & 12.441& 2.030&                &                 &  21 &  31.75\\
3-17 &1026 & 11.805& 2.073&                &                 &  40 &  44.76\\
3-18 & 954 & 13.312& 1.550&                &   25.3          &     &       \\
4-1  & 711 & 12.380& 1.639&                &                 &     &  28.56\\
4-2  & 759 & 11.429& 1.423&                &   21.6$\div$30.1&  35 &  32.53\\
4-16 & 883 & 12.856& 1.546&                &   44.5          &  60 &       \\
\hline
\end{tabular}
\end{table*}
\begin{figure}
\includegraphics[width=\columnwidth]{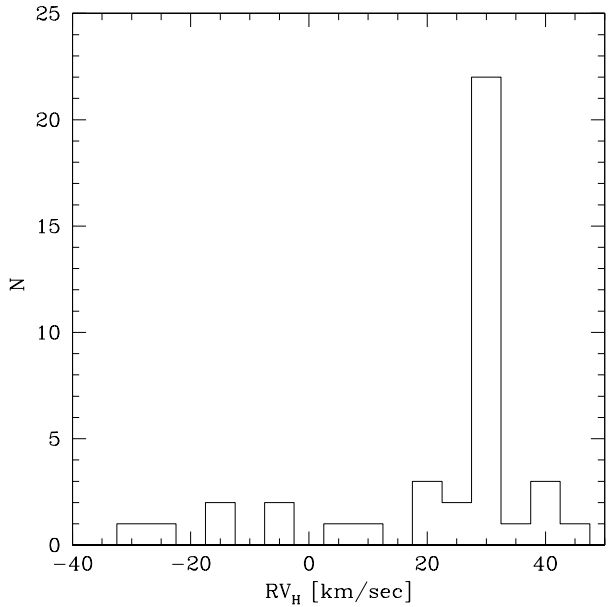}
\caption{Distribution of radial velocities from the present study.}
\end{figure}

\begin{figure}
\includegraphics[width=\columnwidth]{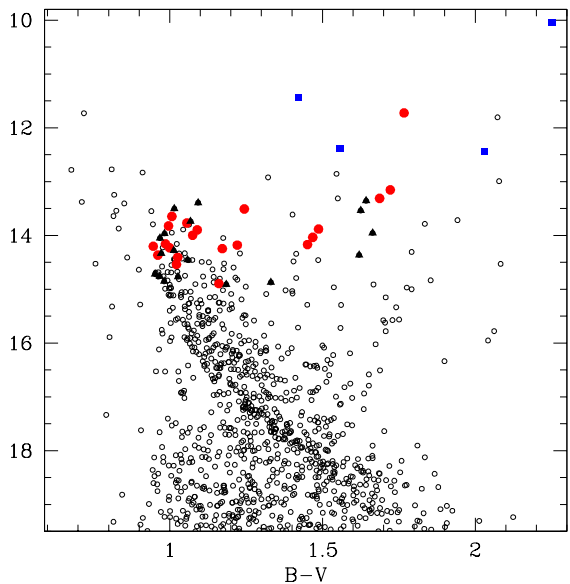}
\caption{CMD of NGC~2112. Filled circles (red when printed in color) indicate radial velocity
members, while filled triangles (black in color) non-members. Filled (color coded in blue)
squares are member stars for which we do not have either photometry or radial velocity.
See text for additional details.}
\end{figure}

\begin{figure}
\includegraphics[width=\columnwidth]{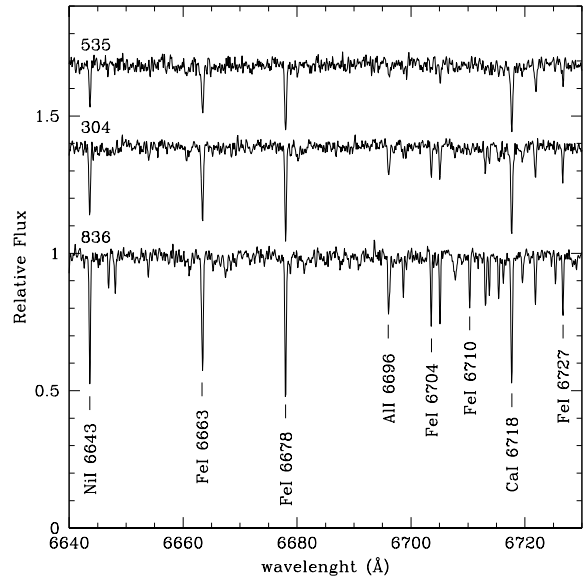}
\caption{Extracted spectra for the 3 MIKE member stars.
A few important lines are indicated}
\label{fig1}
\end{figure}

%
\section{Membership and Cluster Mean Radial Velocity}
%

We derived radial velocities
of the target stars using the IRAF {\it fxcor} task, which
cross-correlates the object spectrum with a template. As a template,
we used a synthetic spectrum calculated by SPECTRUM \footnote{SPECTRUM
is the local thermodynamical equilibrium (LTE) spectral synthesis
program freely distributed by Richard O. Gray. See: http://www.phys.appstate.edu/spectrum/spectrum.html}
with roughly the same atmospheric
parameters and metallicity of the observed stars. The final errors in
the radial velocities, as provided by {\it fxcor}, were typically less than 1.0 km/s
for most of the Hydra stars and less than 0.3 km/s
for MIKE targets (see Table~2).  In the case of Hydra spectra, these
have to be considered the real errors since they have been taken
from the average of the two different exposures and their combined errors. In the
case of MIKE spectra, having only one exposure, we consider the reported
error as a lower limit of the real error.\\
The distribution of radial velocities is shown in the histogram in Fig.~4.
The bulk of stars form a peak in the heliocentric radial velocity
distribution around 30-31 km/s, allowing us to define a mean velocity
for the cluster and the dispersion, $\sigma$.
These turn out to be:

\begin{center}
\begin{equation}
RV_{\rm H}=30.9\pm0.4\ km/s
\end{equation}
\end{center}

\noindent
To derive this value we used 21 stars, which are listed in Table~2
as cluster members.
These stars were selected using an interactive procedure in which we calculated
an initial value for RV$_{\rm H}$ and $\sigma$. Then
stars having radial velocities more than 3 $\sigma$ from the mean were rejected as non-members
and a new RV$_{\rm H}$ and $\sigma$ computed. This procedure
was iterated until no more stars were rejected.\\
It is, however, possible that some of the rejected stars are binary stars.\\
In Table~4, we compare our measurements with literature values.
Radial velocity data for NGC 2112 are poor and very inhomogeneous.
We have 3 stars in common with BWGO96, one with FJ93 and 2 with
MM07. In all cases, the radial velocities are compatible within the errors,
as are the membership assignments.\\
We confirm the result of BWGO96 that star $\#$2-4 (our star 656) is a member.\\
One more star, for which we do not have new radial velocity, can
be considered a member, following MM07, if it is a binary: this is
star  $\#$3-18 (our 954).\\
Finally, we find that $\#$3-16, 4-1 and 4-2
(our stars 1123, 711 and 759) are member stars following the analysis in MM07 and looking at Table~5.\\

\noindent
We make use of the CMD to get additional
information on the cluster membership. In Fig.~5, we indicate
with filled circles (red when printed in color) the radial
velocity members, and with filled triangles (black when printed in color)
the radial velocity non-members.
Additionally, we plot as filled squares (blue in color) stars
that are members, but for which we do not have radial velocity
measurements  (1123, 711 and 759), and star 362 ($\#$3-16), for which we have measured its
radial velocity but not photometry, which we take from R85..

Clearly, members and non-members mix up in a way that, without
radial velocities, it would not be possible to discriminate between them.
Our member stars are partly located in the TO region and partly
trace the sub-giant branch and RGB of the cluster.

Among this membership sample, we find two stars with the radial velocities typical of members
but located far from the most important loci in the CMD.
They are stars $\#$323 at (V,B-V) = (14.896, 1.161) and $\#$782 at  (13.510,1.244). We suggest that these stars may be binary
members, as is star $\#$3-18 (our 954, see MM07), which must be confirmed by future studies
(but see Section.~7 for additional details).\\
\noindent
Unfortunately, we cannot use proper motions (see Table~2) to improve our 
membership assignments due to the large associated errors.

\begin{table}
\caption{Atmospheric parameters of MIKE member stars}
\label{tab1}
\centering
\begin{tabular}{lcccc}
\hline
ID & T$_{\rm eff}$(K) & log(g) & v$_t$(km/s)& [Fe/H]\\
\hline
535&    6650&  3.85&  1.70&  0.15$\pm$0.02 \\
304&    5980&  3.65&  1.00&  0.19$\pm$0.02 \\
836&    5130&  3.48&  1.02&  0.13$\pm$0.01 \\
\hline
\end{tabular}
\end{table}
\section{Abundance measurements}

\subsection{Atomic Parameters and Equivalent Widths}

We performed the analysis of chemical abundances on the 3 members observed with MIKE using the 2007 version of the free available program {\it MOOG} developed by Chris
Sneden\footnote{http://verdi.as.utexas.edu/moog.html}
and using model
atmospheres by Kurucz (1992). {\it MOOG} performs a local thermodynamic
equilibrium (LTE) analysis.
We derived equivalent widths of spectral lines
by Gaussian fitting of spectral features.
Repeated measurements show a
typical error of about 5 m\AA~ for the weakest lines.
The line list was
taken from Carraro et al. (2008). The log(gf)
parameters of these lines were re-determined by a solar-inverse
analysis measuring the equivalent widths from the NOAO solar spectrum (Kurucz et al. 1984), adopting 
standard solar parameters (T$_{\rm eff}$ = 5777 K, log(g) = 4.44, and v$_{\rm t}$ = 0.8 km/s).
The $O$ abundance was obtained from the IR triplet at 7771-5 \AA~,
while the Na abundance was obtained from the spectral doublets at 5662-68 and 6154-60 \AA~.
These features are well known to be affected by NLTE effects.
For this reason we applied NLTE correction to the output
LTE abundances, obtained from Gratton \& al. (1999).

\begin{figure}
\includegraphics[width=\columnwidth]{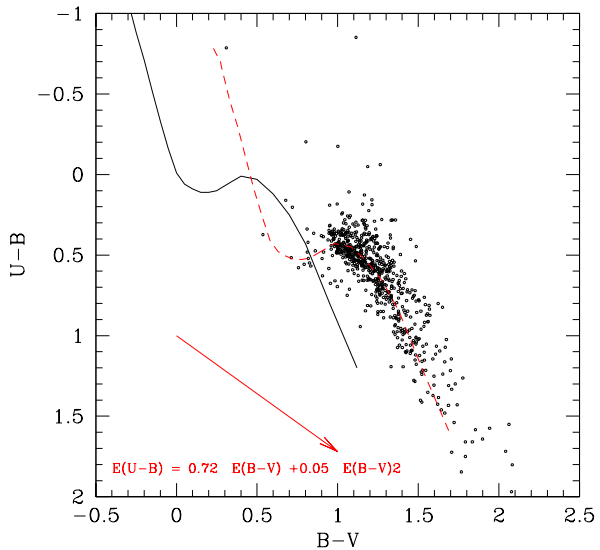}
\caption{Two-color diagram for NGC~2112 stars. The solid line is the Schmidt-Kaler (1982)
empirical ZAMS and the dashed line is the same ZAMS shifted by E(B-V)=0.60
along
the reddening vector (the arrow) for a normal reddening law.}
\label{fig1}
\end{figure}

\subsection{Atmospheric Parameters}
Initial estimates of the atmospheric parameter T$_{\rm eff}$ were
obtained from photometric observations using the relations from Alonso et al. (1999).
We adopted E(B-V) values from Carraro et al. (2002) to correct colours for interstellar extinction.
We then adjusted the effective temperature by minimizing the slope of
the abundances obtained from Fe I lines with respect to the excitation
potential in the curve of growth analysis.
Initial guesses for the gravity log(g) were derived from the canonical
formula:

\begin{center}
\begin{equation}
\log\left(\frac{g}{g_{\odot}}\right) =
\log\left(\frac{M}{M_{\odot}}\right) + 4
\log\left(\frac{T_{\rm{eff}}}{T_{\odot}}\right)
- \log\left(\frac{L}{L_{\odot}}\right)
\end{equation}
\end{center}

\noindent
In this equation, the mass $M/M_{\odot}$ was derived from the
comparison between the position of the star in the Hertzsprung-Russell
diagram and the Padova isochrones (Girardi et al. 2000).
The luminosity $L/L_{\odot}$ was derived from the absolute magnitude
$M_V$,  assuming a distance moduli of
$(m-M)_V=11.6$.
The bolometric correction
(BC) was derived from the BC-Teff relation from Alonso et al. (1999).
The input log(g) values were then adjusted in order to satisfy the
ionization equilibrium of Fe I and Fe II during the abundance
analysis.
Finally, the microturbulence velocity is given by the relation (Houdashelt et al.\ 2000):

\begin{center}
\begin{equation}
v_{\rm t}=2.22-0.322\times log(g)
\end{equation}
\end{center}

We then adjusted the micro-turbulence velocity by minimizing the slope
of the abundances obtained from Fe I lines with respect to the
equivalent width in the curve of growth analysis.
The adopted values for all these parameters, together with {\rm [Fe/H]}, are reported
in Table~6.  The results of the abundance analysis are listed
in Table~7, where the abundances of the main elements 
are reported with their uncertainties. Two stars turn out to be giants ($\#$ 304 and 836) , while star 
$\#535$ is clearly a dwarf. This explains the different number of lines
used in the determination of the different elemental abundances.\\
As a final remark, we also performed an abundance analysis for the two radial
velocity non-members, and found that the two stars $\#$261 and $\#$717 have {\rm [Fe/H]} =
+0.28$\pm$0.02 and +0.30$\pm$0.03, confirming their nature as non member stars.\\

\noindent
Examples of our extracted spectra are illustrated in Fig.~6, in which the
spectrum of 3 MIKE member stars are shown and some interesting lines indicated.\\

\noindent
The mean metallicity we derive ({\rm [Fe/H]}=+0.16$\pm$0.03) is significantly different
from any previous spectroscopic estimate. The closest determination is the one by BWGO96,
who found {\rm [Fe/H]}=-0.15$\pm$0.15. Our result rules out any possibility that NGC~2112
is very metal poor, as suggested in early studies. It would have been very
unusual to have such a metal poor cluster in the solar neighborhood.
Our result, in fact, suggests that NGC~2112 has a typical solar vicinity metal
abundance, being as metal rich as the Hyades (Boesgaard \& Friel 1990).\\

\begin{figure*}
\includegraphics[]{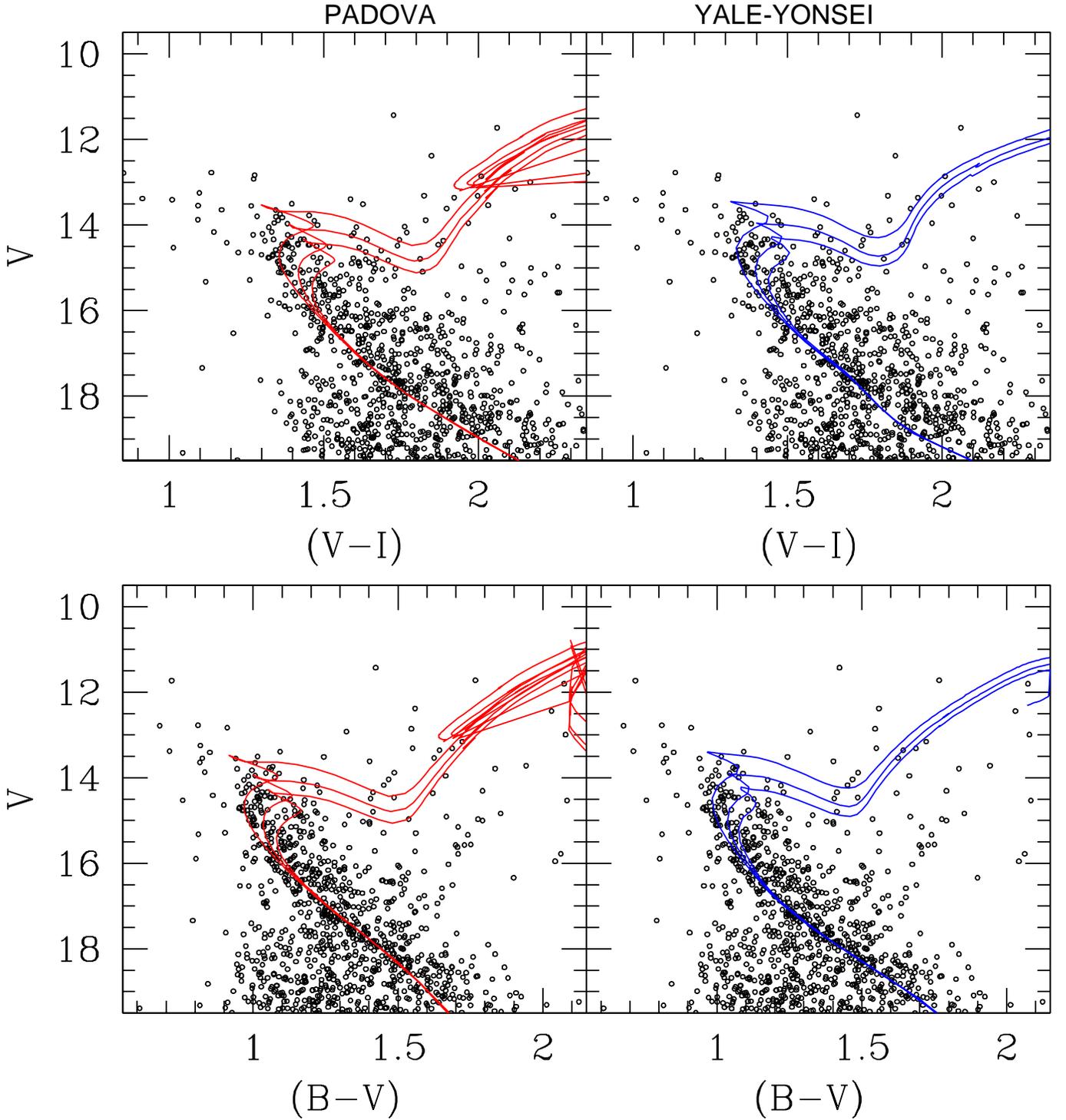}
\caption{Isochrone fits for the derived metallicity and varying ages.
Bottom panels show the fits in the V vs (B-V) plane, upper panels in the V vs (V-I) planes.  Left panel refers to Padova isochrones, and right 
panels to Yale-Yonsei isochrones.}
\label{fig3}
\end{figure*}


\begin{table*}
\caption{Abundance analysis from MIKE cluster members. After each value, in parenthesis,
the number of lines $N$ used is indicated. Values derived from just one line do not have
any error associated.}
\fontsize{8} {8pt}\selectfont
\begin{tabular}{lcccccc}
\hline
\multicolumn{1}{c} {$Element$} &
\multicolumn{1}{c} {$\#535$}  &
\multicolumn{1}{c} {$N$}  &
\multicolumn{1}{c} {$\#304$}  &
\multicolumn{1}{c} {$N$}  &
\multicolumn{1}{c} {$\#836$}  &
\multicolumn{1}{c} {$N$}  \\
\hline
$\rm{[Fe/H]}$ &  0.15$\pm$0.02 &(19) &  0.19$\pm$0.02 &(71) & 0.13$\pm$0.01 &(107) \\
$\rm{[O/H]_{LTE}}$  &  0.45$\pm$0.07 &(3)  &  0.36$\pm$0.01 &(3)  & 0.27$\pm$0.03 & (3)  \\
$\rm{[O/H]_{NLTE}}$  &  0.03$\pm$0.07 &(3)  &  0.17$\pm$0.01 &(3)  & 0.23$\pm$0.03 & (3)  \\
$\rm{[Na/H]_{LTE}}$ &  0.35$\pm$0.13 &(2)  &  0.20$\pm$0.06 &(4)  & 0.30$\pm$0.06 & (3)  \\
$\rm{[Na/H]_{NLTE}}$ &  0.24$\pm$0.13 &(2)  &  0.15$\pm$0.06 &(4)  & 0.23$\pm$0.06 & (3)  \\
$\rm{[Mg/H]}$ &  0.13          &(1)  &  0.18          &(1)  & 0.27$\pm$0.03 & (2)  \\
$\rm{[Al/H]}$ &                &     &  0.10$\pm$0.09 &(2)  & 0.05$\pm$0.03 & (3)  \\
$\rm{[Si/H]}$ &  0.33$\pm$0.02 &(3)  &  0.11$\pm$0.06 &(5)  & 0.11$\pm$0.05 & (8)  \\
$\rm{[Ca/H]}$ &  0.18$\pm$0.07 &(7)  &  0.23$\pm$0.04 &(13) & 0.07$\pm$0.05 &(13)  \\
$\rm{[Ti/H]}$ &  0.20$\pm$0.04 &(2)  &  0.43$\pm$0.09 &(7)  & 0.23$\pm$0.04 &(23)  \\
$\rm{[V/H]}$  &                &     & -0.01          &(1)  & 0.31$\pm$0.06 &(10)  \\
$\rm{[Cr/H]}$ &  0.51$\pm$0.05 &(2)  &  0.41$\pm$0.08 &(6)  & 0.14$\pm$0.03 &(11)  \\
$\rm{[Mn/H]}$ & -0.18          &(1)  &  0.03$\pm$0.03 &(3)  & 0.10$\pm$0.02 & (3)  \\
$\rm{[Co/H]}$ &                &     &  0.27$\pm$0.07 &(2)  & 0.24$\pm$0.06 & (4)  \\
$\rm{[Ni/H]}$ &  0.17          &(1)  &  0.24$\pm$0.05 &(15) & 0.21$\pm$0.03 &(29)  \\
$\rm{[Cu/H]}$ &                &     &  0.38          &(1)  & 0.12          & (1)  \\
$\rm{[Y/H]}$  &                &     &  0.45          &(1)  & 0.51$\pm$0.24 & (2)  \\
$\rm{[Ba/H]}$ &  0.59          &(1)  &  0.84$\pm$0.06 &(2)  & 0.40$\pm$0.01 & (2)  \\
\hline
\end{tabular}
\end{table*}

\noindent
Karatas \& Schuster (2006) provided a new calibration of the relation between the metallicity
of a dwarf star and $\delta_{0.6}$, namely the (U-B) excess/deficiency with respect
to the Haydes sequence at (B-V)$_{o}$ = 0.6. We compared the distribution of dwarf
stars in NGC~2112 at (B-V)$_{o}$ $\sim$ 0.6 with respect to a ZAMS from Girardi et al. (2002)
having the same metallicity of the Hyades ([Fe/H]=0.17). We found that the useful stars
(13 in number) have $\delta_{0.6}$=-0.02$\pm$0.11. Despite the scatter,
this  $\delta_{0.6}$ implies a metallicity close to our spectroscopic determination (see Table~3).

\section{Abundance ratios}
We derived abundance ratios for the 3 MIKE member stars listed in Table~7.
Previously, only BWGO96 provided an estimate of a few abundance
ratios in NGC~2112, but based on just one star.
At any rate, we compare our findings with BWGO96 (their Table~6),
and find basic agreement with [O/Fe] and [Na/Fe]. However, their [Al/Fe]
is much larger than our value.\\
\noindent
Being close to the Sun and having roughly the same metallicity,
it is useful to make a detailed comparison of the 
chemical properties of NGC~2112 with the solar neighborhood stars and star clusters.

\subsection{Comparison with field stars}
Bensby et al. (2005) present 
a detailed abundance analysis for a sample of 102 F and G dwarf
stars in the solar vicinity.\\

\noindent
{\bf $\alpha-$ elements}\\
As shown by Bensby et al (2005), $O$, $Mg$, $Si$, $Ca$ and $Ti$ exhibit similar trends
in the Galactic thin disk. At the Fe abundance of NGC~2112 (+0.16) , these ratios
are in the range of -0.15$:$0.00, 0.00$:$+0.15, 0.00$:$+0.15, 0.00$:$+0.15,
and -0.05$:$0.05 dex, respectively.\\
According to our findings, abundance ratios for these five elements in NGC~2112 are in agreement
with the thin disk values within the errors . This confirms that NGC~2112 is a typical
thin disk star cluster. 
The overall ${\rm [\alpha/Fe]}$ ratio turns out to be
0.04$\pm$0.03.\\

\noindent
{\bf Iron peak elements}\\
We can compare  only $Ni$ and $Cr$ with Bensby et al (2005). 
Ni is basically in agreement with the thin disk trends, 
whereas the {\rm [Cr/Fe]} ratio is marginally overabundant.\\

\noindent
{\bf Al and Na}\\
While the $Na$ abundance relative to $Fe$ is consistent with thin disk values, we find that
Al is significantly under-abundant. \\

\noindent
{\bf r- and s-process elements}\\
We measured $Y$ and $Ba$ abundance ratios. While the $Ba$ abundance in NGC~2112
is consistent with the thin disk trend, we find that the Y abundance is significantly
larger that typical thin disk values.\\

\subsection{Comparison with open clusters}
In general, there is not much information on abundance ratios in open clusters,
and only in the last few years efforts have been done to improve this
situation.\\
Here, we compare our NGC~2112 abundance ratios with the results presented in Friel et al. (2003).
They provide a detailed abundance
analysis of the old star cluster Collinder~261
and compare its abundance ratios with a sample of 10 open clusters (see their Table 7).
From this table, we extract estimates for NGC~2360 and NGC~6819, two nearby clusters
having roughly the same ages and metal abundances as NGC~2112.
We find that within the errors, NGC~2360 and NGC~6819 possess the same {\rm [$\alpha$/Fe]} as NGC~2112,
+0.03 and 0.00, respectively.\\
The $Na$ abundance of NGC~2112 is similar to NGC~2360 but significantly lower than in NGC~6819.
As for $Al$, we can make a comparison only with NGC~6819, for which {\rm [Al/Fe]}
is similar to the value we determine for NGC~2112.\\
Unfortunately, neither information about the Iron-peak elements nor for s- and r- process 
elements are reported in Friel et al. (2003).\\

\noindent
Overall, with a few exceptions, we find that NGC~2112 is a genuine
thin disk population cluster.

\section{Cluster fundamental parameters}
Having an estimate of the metal content ({\rm [Fe/H]}=+0.16),
and of the $\alpha-$ element abundances (${\rm [\alpha/Fe]}$=+0.04),
we are now in the position to derive more reliable estimates
of the cluster parameters.\\
The reddening value in the direction of NGC 2112 predicted by Schlegel et al. (1998) maps
is  E(B-V) = 1.01. This has to be considered as an upper limit to the reddening
since it takes into accounts the absorption all the way to infinity.\\
To get an independent estimate of the reddening in the direction of NGC~2112, we make use
of the two-color diagram (TCD) in Fig.~7  since
we provide deep $U$ band photometry for the first time.
Here, the solid line is the zero reddening
empirical Zero Age Main Sequence (ZAMS) from Schmidt-Kaler (1982). The same ZAMS,
shifted by E(B-V)=0.60 is shown as a dashed line. The shift has been performed
adopting the standard reddening law (see the expression in the bottom of Fig.~7),
and the reddening vector is indicated with a solid arrow.\\

\begin{figure*}
\includegraphics[]{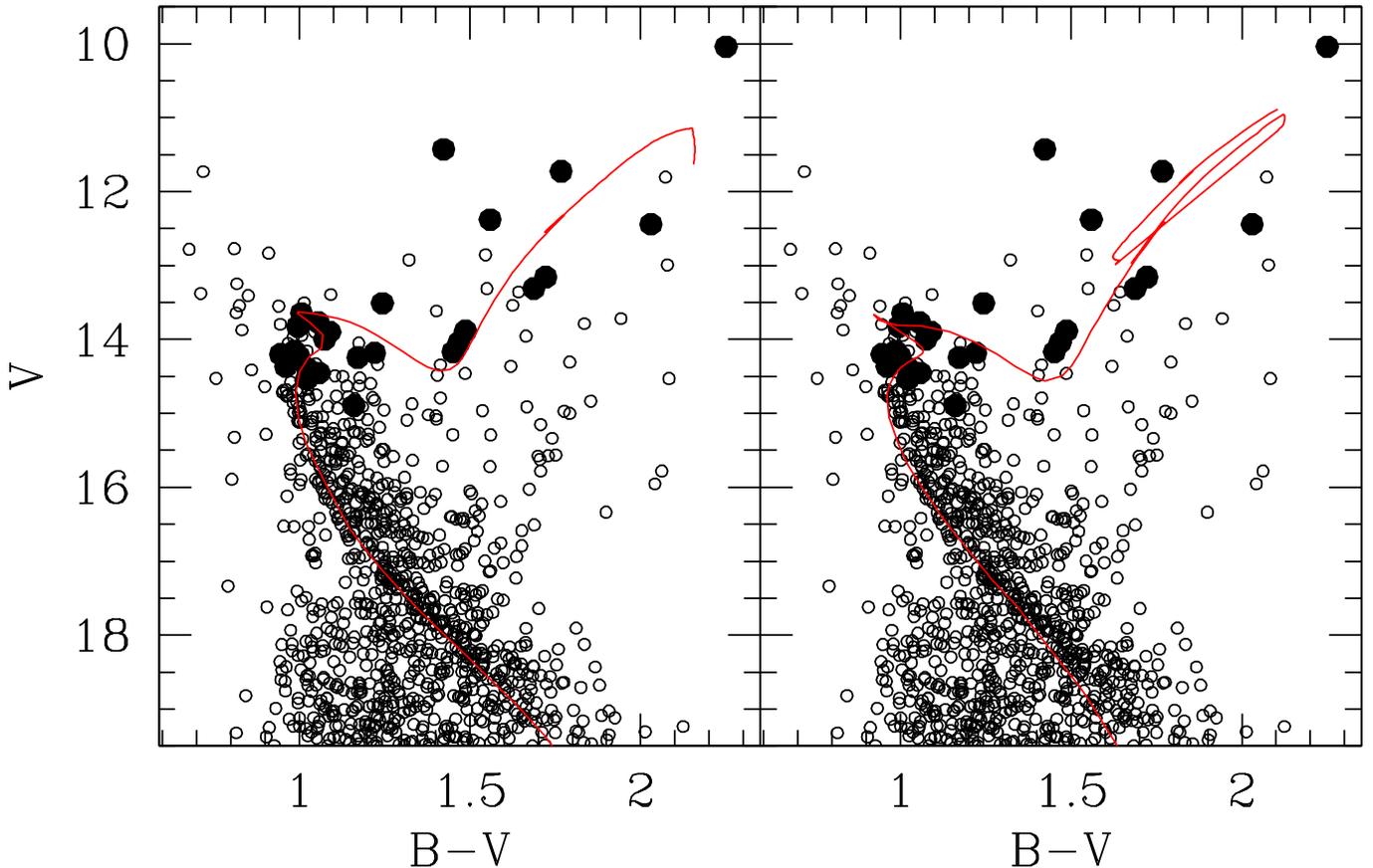}
\caption{The best fit isochrone is super-imposed on the NGC~2112 CMD where radial velocity members
are indicated with solid circles. Left panel refers to Yale-Yonsei isochrones, and the right
panel to Padova isochrones. The isochrones have been adjusted using the values
listed in Table~8.}
\label{fig3}
\end{figure*}

\noindent
We compare the distribution of stars in the various color combination  CMDs and sets of theoretical
Padova (Girardi et al. 2000)  and Yale (Demarque et al. 2004, Kim et al. 2002, Yi et al 2001) isochrones.
The physical ingredients of the two sets, and the possible sources
of different results, have been discussed exhaustively in Carraro et al. (2006),
where a similar exercise has been done for the very metal rich
open cluster NGC~6791.\\
In Fig. 8 we superimpose isochrones with [Fe/H]=+0.16 and ${\rm
[\alpha/Fe]}$=+0.04 
for 3  ages (1.5, 2.0, and 2.5  Gyrs).
The fitting is performed in the left panels for Padova models,
and in the right panels for Yale models.

\noindent
To derive more constrained basic parameters, we consider in Fig.~9
the star distribution in the V vs (B-V) plane, and we highlight the
radial velocity members using filled circles. The best fit isochrone is found
for an age of 1.8$\pm$0.3 Gyr, where the associate error has been derived by
trying different age isochrones.\\
We find that the both the Yale-Yonsei and Padova sets 
fit the star distribution well, and we summarize
the derived values for the basic parameters in Table~8.\\

\noindent
On average, we obtained (m-M)$_{V}$=11.75$\pm$0.15 and E(B-V)=0.60$\pm$0.10.
The errors here reported have been derived by displacing the best fit isochrone
back and forth in the distance modulus and reddening directions and exploring
the values of distance modulus and reddening which produce acceptable fits. 

The two sets of models imply the same values for NGC~2112 basic
parameters within the errors. It is worth noticing, however, that the fit
implies a sizeable difference in the mean E(B-V), in the sense that the reddening 
that inferred using Yale-Yonsei models is larger than that inferred
from Padova models. At the same age and metal abundance, the apparent Yale-Yonsei RGB is bluer than the Padova one, resulting in 
a larger reddening when fitting the observed RGB.\\

\noindent
Using the
heliocentric rectangular Galactic coordinates X= 9.3 kpc, Y= -400 pc, and Z =  -200 pc and 
assuming the Sun distance from the Galactic center is 8.5 kpc, NGC 2112 is located 940$\pm$70 pc from the Sun towards the anti-center direction. 
Consequently, we infer a distance of 9.3 kpc from the Galactic center.\\

\begin{table}
\caption{Abundance ratios from MIKE cluster members}
\fontsize{8} {8pt}\selectfont
\begin{tabular}{lccc}
\hline
\multicolumn{1}{c} {$Element$} &
\multicolumn{1}{c} {$\#535$}  &
\multicolumn{1}{c} {$\#304$}  &
\multicolumn{1}{c} {$\#836$}  \\
\hline
$\rm{[O/Fe]_{NLTE}}$  &  -0.12  &  -0.02  &  0.10\\
$\rm{[Na/Fe]_{[NLTE]}}$ &  0.09  &  -0.04  &  0.10\\
$\rm{[Mg/Fe]}$ & -0.02  & -0.01  &  0.14\\
$\rm{[Al/Fe]}$ &        & -0.09  & -0.08\\
$\rm{[Si/Fe]}$ &  0.18  & -0.08  & -0.02\\
$\rm{[Ca/Fe]}$ &  0.03  &  0.04  & -0.06\\
$\rm{[Ti/Fe]}$ &  0.05  &  0.24  &  0.10\\
$\rm{[V/Fe]}$  &        & -0.20  &  0.18\\
$\rm{[Cr/Fe]}$ &  0.36  &  0.22  &  0.01\\
$\rm{[Mn/Fe]}$ & -0.33  & -0.16  & -0.03\\
$\rm{[Co/Fe]}$ &        &  0.08  &  0.11\\
$\rm{[Ni/Fe]}$ &  0.02  &  0.05  &  0.08\\
$\rm{[Cu/Fe]}$ &        &  0.19  & -0.01\\
$\rm{[Y/Fe]}$  &        &  0.26  &  0.38\\
$\rm{[Ba/Fe]}$ &  0.44  &  0.65  &  0.27\\
\hline
\end{tabular}
\end{table}

\begin{table}
\caption{Summary of NGC 2112 fundamental parameters derived from the comparison
of different isochrone sets.}
\label{tab1}
\centering
\begin{tabular}{lccc}
\hline
Models & Age & E(B-V) & (m-M)$_V$\\
\hline
Yale-Yonsei & 1.8$\pm$0.3 & 0.63$\pm$0.05  & 11.80$\pm$0.10 \\
Padova      & 1.8$\pm$0.3 & 0.57$\pm$0.05  & 11.75$\pm$0.10 \\
\hline
\end{tabular}
\end{table}
\begin{figure}
\includegraphics[width=\columnwidth]{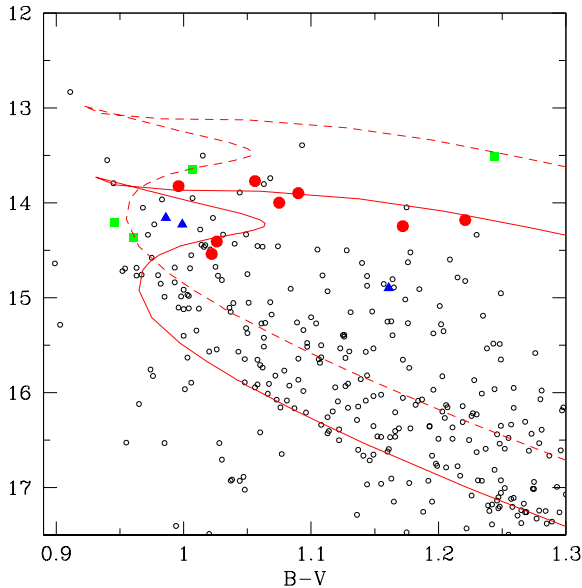}
\caption{A zoom of the TO region in NGC 2112 CMD. Filled circles indicate {\it bona fide}
single stars, filled squares possible equal mass binary systems, and filled triangles
multiple systems or unequal mass binaries. The solid line is the best fit isochrone
which is also shown in Fig.~9, whereas the dashed line is the same isochrone shifted
upward by 0.75 mag. to mimic the location of equal mass binaries.}
\label{fig3}
\end{figure}

\section{A zoom of  the TO region: getting additional clues to the binary population}
Now that we have determined the cluster's fundamental parameters, we can use the 
best fit isochrone as a tool to investigate possible binaries among
cluster members from a purely photometric point of view.
In Fig~10 we provide a zoom of the TO region, and indicated with filled circles (color
coded in red) radial velocity members.
The solid line is the best fit Padova isochrone (see Fig.~9). 
The dashed curve is the same isochrone, but shifted up by 0.75 mag, to illustrate
the locus of unresolved equal mass binaries.
This exercise is suggesting us that several stars that lie off the best fit isochrone
may in reality be unresolved binaries. In fact some of them lie very close to the binary
locus, and we indicate them as filled squares (color coded in green) in the CMD
of Fig.~10. They are stars $\#$707, 824, 782 and 601.
At the same time, we indicated as filled triangles (blue when printed in color)
stars which lie neither in the single star nor in the binary star sequence.
These are stars $\#$714, 707, and 323. One may speculate that these are unequal mass
binaries or maybe systems with more than two components.
If we refer only to the TO region, we are left with 14 single member stars and 7 
possibly multiple systems, which implies a rough binary percentage of 33$\%$.

\section{Discussion and conclusions}
In this paper we presented new photometric and spectroscopic data
in the field of the old open cluster NGC~2112.
This new dataset allowed us to revise the cluster's fundamental parameters
and clarify a long lasting debate on its properties, which for many years have been
poorly constrained due to the high level of field star contamination.\\
By means of multi-fiber spectroscopy, we measured radial velocity for
40 stars and found 21 radial velocity members. This, in turn, allowed
us to clean the CMD, providing a better comparison with stellar models.\\

\noindent
The most important result of our study is that the cluster has a metallicity
much higher than previous determinations, and somewhat higher than the Sun.
Also the $\alpha-$ elements are marginally enhanced with respect to the Sun,
but still compatible with the trends of thin disk stars in the solar
vicinity, as are all the other elements we measured.\\
Therefore NGC~2112 is typical of old, thin disk star clusters, as metal rich as the Hyades (Boesgaard \& Friel 1990),  and located at less than 1 kpc from the Sun in the anticenter direction.\\
In addition, we confirmed the age of the cluster ($\sim$ 1.8 Gyr), previously derived on a purely
photometric basis (Carraro et al. 2002). As for the distance and reddening, our
values are in agreement within the errors of previous determinations.\\

\noindent
This study stresses the importance of performing detailed
membership analysis in Galactic open clusters in order to derive more
robust estimates of their fundamental parameters.

\section*{Acknowledgments}
This research was part of a joint project between Universidad the Chile and Yale University,
partially funded by the Fundaci\'on Andes.
Hydra observations were performed remotely
from Yale by MVM.
CMB acknowledges Universidad de Chile graduate fellowship support from program MECE
Educaci\'on Superior UCH0118 and Fundaci\'on Andes C-13798.
MVM was supported by an NSF Astronomy and Astrophysics Postdoctoral Fellowship,
under award AST04-011640.
GC thanks Doug Geisler, Tom Richtler and Roberto
Barbon for long and interesting discussions on NGC 2112, and the anonymous referee
who helped us to improve the paper presentation.
This study made use of SIMBAD and WEBDA.

%


\begin{thebibliography}{99}
\bibitem[1999]{A99} Alonso, A., Arribas, S. \&
  Mart\'inez-Roger, C. 1999, A\&AS, 140, 261
\bibitem[2005]{ben05} Bensby, T., Feltzing, S., Lundstr\"om, I.,
  Ilyin, I., 2005, A\&A 433, 185
\bibitem[1990]{bp90} Boesgaard, A.M., Friel, E.D., 1990, ApJ 351, 467
\bibitem[1996]{B96} Brown J.A.,
  Wallerstein G., Geisler D., \& Oke J. B. 1996, AJ, 112, 1551
\bibitem[1998]{C98} Carraro, G., Ng, Y.K., \& Portinari,
  L. 1998, MNRAS, 296, 1045
\bibitem[2002]{C02} Carraro, G., Barbon, R., \& Boschetti,
  C.S. 2002, MNRAS, 336, 259
\bibitem[2006]{C06} Carraro G.,
  Villanova, S., Demarque, P., McSwain, M.V., Piotto, G., Bedin, L.R.,
  2006, ApJ 643, 1151
\bibitem[2008]{C08} Carraro G.,
  Geisler, D., Villanova, S., Frinchaboy, P.M., Majewski, S.R., 2008, A\&A 476, 217
\bibitem[2004]{de04} Demarque P.,Wo,Y-H., Kim, Y-C.,
Yi, S.K. 2004, ApJS 115, 667
\bibitem[2002]{dias02} Dias, W.S., Alessi, B.S., Moitinho, A.,
Lepine, J.R.D. 2002, A\&A 389, 871
\bibitem[1993]{F093} Friel E.D., Janes
 K.A., 1993, A\&A 267, 75
\bibitem[2003]{fri02} Friel, E.D., Jacobson, H.R.,
Barrett, E., Fullton, L., Balachendran, S.C., Pilachowski, C.A, 2003, AJ 126, 2372
\bibitem[1987]{G87} Geisler D. 1987, AJ, 94,
  84
\bibitem[1991]{GCM91} Geisler D.,
  Claria J.J., Minniti. D 1991, AJ, 102, 1836
\bibitem[2000]{Gi00} Girardi L.,
  Bressan A., Bertelli G., \& Chiosi, C. 2000, A\&AS, 141, 371
\bibitem[2003]{Gr03} Gratton R. G.,
  Carretta E., Eriksson K., \& Gustafsson B. 1999, A\&A, 350, 955
\bibitem[2000]{H00} Houdashelt M.L., Bell R.A., \&
  Sweigart A.V. 2000, AJ, 119, 1448
\bibitem[1986]{Hor} Horne, K., 1986, PASP 98, 60
\bibitem[2006]{kar06}Karatas Y., Schuster W.J., 2006, MNRAS 371, 1793
\bibitem[2002]{ki02} Kim, Y.-C., Demarque, P., Alexander, D.R.,
2002, ApJS 143, 499
\bibitem[1984]{Ku84} Kurucz R.L.,
  Furenlid I., Brault J., \& Testerman, Larry 1984, sfat.book, K
\bibitem[1992]{K92} Kurucz R.L. 1992, IAUS, 149, 225
\bibitem[1994]{Ja94} Janes, K.A., Phelps, R.L., 1994, AJ 108, 1773
\bibitem[1992]{land} Landolt, A.U., 1992, AJ 104, 340
\bibitem[2007]{Me07} Mermilliod, J.-C., Mayor, M., 2007, A\&A 470, 919
\bibitem[1985]{R85} Richtler T. 1985, A\&AS, 59, 491
\bibitem[1989]{RK89} Richtler T.,
  \& Kaluzny J. 1989, A\&AS, 81, 225
\bibitem[1998 ]{Sc98} Schlegel
 D.J., Finkbeiner D.P., \& Davis M. 1998, ApJ, 500, 525
\bibitem[1982]{sc82}Schmidt-Kaler, Th., 1982, in Schaifers K.,
Voigt, H.H., eds., Numerical Data and Functional Relationships in Science and Technology,
Landolt-B\"orbstein, New Series, Group VI, Vol 2(b), Springer, Berlin, p. 14
\bibitem[1987]{stet}Stetson, P.,B., 1987, PASP 99, 191
\bibitem[2004]{zac} Zacharias, N., Urban, S.E., 
Zacharias, M.I., Wycoff, G.L., Hall, D.M., Monet, D.G., Rafferty, T.J., 2004, AJ 127, 3043
\bibitem[2002]{de04}Yi, S.K, Demarque P.,Kim, Y-C.,
Lee, Y.-W, Ree, C.H., Lejeune, Th., Barnes, S., 2001, ApJS 136, 417

\end{thebibliography}
\end{document}